\documentclass{article}

\usepackage{amsmath,amsfonts,amssymb,amsthm}
\usepackage[margin=1in]{geometry}
\usepackage[colorlinks]{hyperref}
\usepackage{epsfig}
\usepackage{amsmath}
\usepackage{setspace}
\usepackage{graphicx}

\usepackage{subfig}	
\usepackage{xcolor}
\usepackage[utf8]{inputenc}
\usepackage[english]{babel}
\usepackage{amsmath,amssymb,amsfonts}
\usepackage{bm}
\usepackage{textcomp}
\usepackage[T1]{fontenc}
\usepackage{graphicx}
\usepackage{float}
\usepackage{geometry}
\usepackage{color}
\usepackage[Conny]{fncychap}
\usepackage{graphics}
\usepackage{epsfig}
\usepackage{amsmath}
\usepackage{color}
\usepackage{setspace}
\usepackage{graphicx}
\usepackage{subfig}
\usepackage[colorinlistoftodos]{todonotes}

\usepackage{times}
\usepackage{graphicx}
\usepackage{epstopdf} 
\usepackage{authblk}

\usepackage[english]{babel}
\usepackage{physics}

\usepackage{amsmath}
\usepackage{graphicx}
\usepackage[noadjust]{cite}

\begin{document}
\title{InAs three quantum dots as working substance for quantum heat engine}
\author[1 2]{H. Ait Mansour}
\author[1]{F. El Ayachi}
\author[2]{M. Faqir}
\author[1]{M. El Baz}
\affil[1]{ESMaR, Faculty of Sciences, Mohammed V University in Rabat, Morocco.}
\affil[2]{LERMA Laboratory, Aerospace Engineering School, International University of Rabat, Morocco}
\vspace{-0.6cm} 
\maketitle
\vspace{0.5cm}

\begin{abstract}
Heat engines are considered a valuable resource for modern society. The development of these systems leads to the production of heat engines with high efficiency despite their small size, called quantum heat engines. Among these, the quantum Otto cycle which is considered a fundamental thermodynamic cycle in classical heat engines, has also found applications in the realm of quantum heat engines. In this paper, we consider three InAs quantum dots as a working substance, which allows the engine to operate at very small scales, in the presence of an electric field, and the Forster mechanism, which describes the transfer of energy between quantum dots and affects thus the engine's behavior. In this regard, we study the behavior of the work performed by the engine and the entanglement in the system as the Forster parameter is varied. We find that the work performed by the engine is affected by the Forster interaction and the electric field and that the entanglement in the system also changed as the Forster parameter was changed. Finally, we study the influence of entanglement on the work performed by the engine. We find and discuss the intricate relation between the entanglement and the performance of the engine.
\end{abstract}
\section{Introduction}
In the last decade, an increasing number of studies have focused on the realization and development of technologies at micro, nano, and atomic scales, using the quantum phenomena with applications in several areas, especially quantum information processing and quantum thermodynamics \cite{kosloff2013quantum}, and quantum technologies. Among these applications, heat engines are considered a valuable resource for modern society. The development of these engines leads to the production realization of heat engines with high efficiency despite their small size, called quantum heat engines \cite{RevModPhys1,PhysRevLett2,PhysRevE3,PhysRevA4,PhysRevE5,PhysRevE6,PhysRevLett7,PhysRevE8,PhysRevE9,PhysRevE10}. In a quantum heat engine, one can produce work from the heat flow of the system between the hot and the cold thermal bath. Quantum mechanics can describe the operating mechanism of the engine and quantum thermodynamics laws using the quantum version of Carnot, Otto, Stirling, and Diesel Cycles \cite{shaw2008comparing,feldmann1996heat,feldmann2004characteristics,rezek2006irreversible}. The choice of the thermodynamic cycle and the working substance is very important to increase the efficiency and the work performed by the engine.\newline 
In this paper, we propose three InAs quantum dots in the presence of an external electric field as a working substance of a quantum heat engine. The motivation for this choice is the possibility of realizing it experimentally, and also its high electron mobility \cite{dayeh2007high,chang2013inas}, which means that electrons can move quickly and efficiently through the material. This is an important feature in nano-scale devices as it allows for fast and accurate controlling. InAs also have a direct band gap \cite{massidda1990structural}, which means that they can efficiently absorb and emit light. In addition InAs double quantum dots exhibit a strong Coulomb blockade effect\cite{mittag2021few}, in which the flow of electrons through the quantum dots is strongly suppressed at low temperatures. This can be used to control the flow of electrons to perform quantum operations. Also, the energy levels of InAs can be precisely controlled using electrical gates, allowing for precise manipulation of the quantum system. Since InAs can be easily integrated with other materials, such as GaAs and Si, which are commonly used in electronics, this makes it possible to incorporate InAs quantum dots into a variety of devices.\\
In this work, we assume that the engine operates following a quantum Otto cycle \cite{mozurkewich1982optimal,thomas2011coupled,petrescu2016otto}, this latter is a thermodynamic process that converts heat into useful work, just like a classical Otto cycle used in the internal combustion engine. However, the quantum nature of the system allows for the possibility of more efficient and powerful engines. Our objective is to examine the work performed by the heat engine by changing the external electric field in the adiabatic stages, and to study the engine's behavior against the Forster mechanism effects \cite{forster}. We also investigate the effect of the temperature of the baths on the efficiency of the quantum heat engine. 

\section{MODEL AND THE QUANTUM OTTO ENGINE }
\subsection{Working substance and thermalization}
 We propose a quantum system composed of three coupled semiconductors InAs quantum dots with a small space between them \cite{mansour2022wigner,nishiba}, as a working substance for the quantum Otto engine (see Figure 1). We use the excitonic electric dipole moments as a qubit in each quantum dot, under an external electric field $\vec{E}$, and rely on the Forster mechanism between the excitons to explain the energy transfer between the qubits \cite{forster}.
 \begin{figure}[H]
	\centering
	\subfloat{\includegraphics[width=0.5\textwidth]{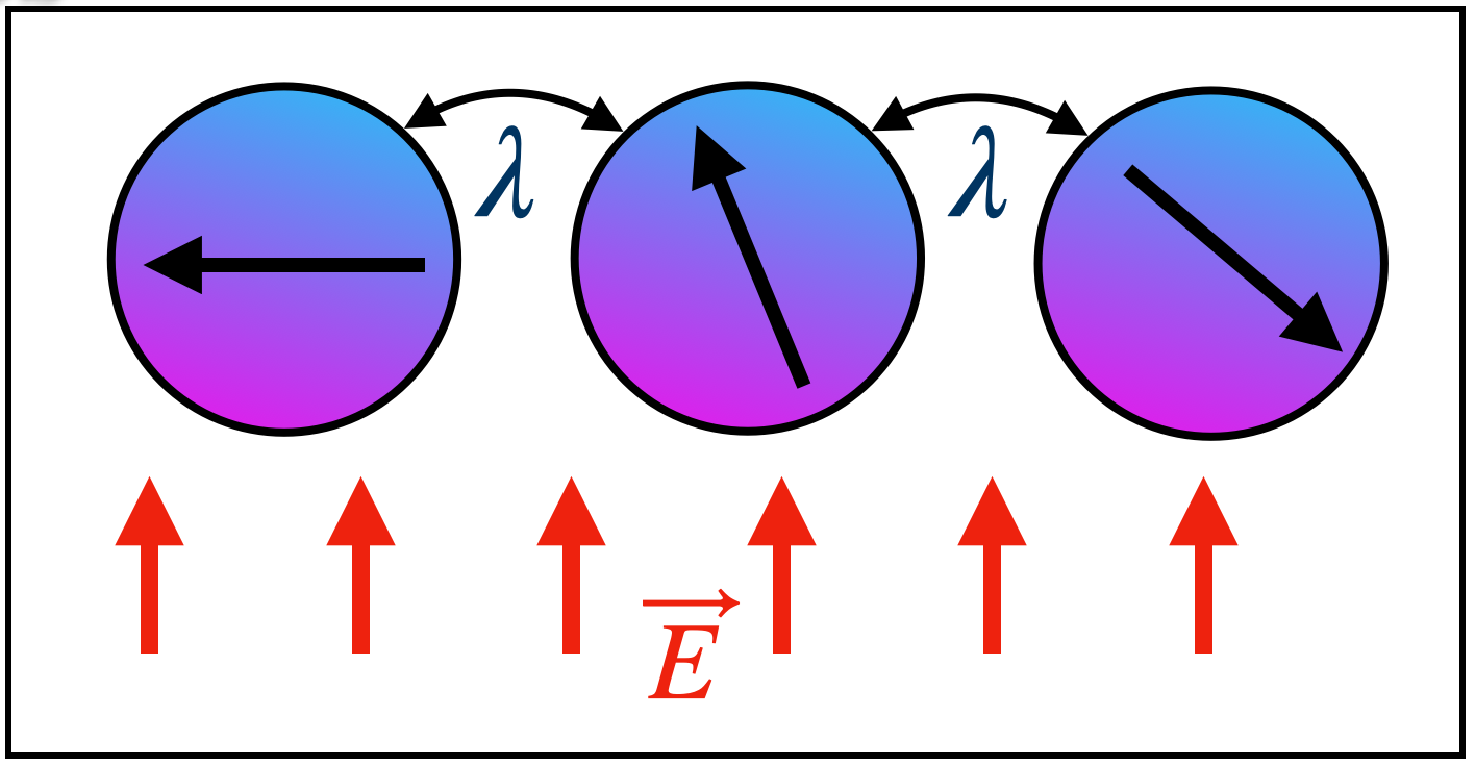}}
 \caption{\label{fii}} Schematic of the three quantum dots.
\end{figure}
 The Hamiltonian of the system, when an electric field $\Vec{E}$ is applied, is given by,
\begin{equation}
\label{eq1}
H =  \sum_{i=1}^{3} \hbar\omega_i [S_z^i +\dfrac{1}{2}] +  \sum_{i=1}^{3} \hbar\Omega_i S_z^i +  \sum_{\substack{i,j=1\\i\ne j}}^{3} \hbar J_z [S_z^i +\frac{1}{2}][S_z^j+\frac{1}{2}]+
\dfrac{1}{2}
 \sum_{\substack{i,j=1\\i \ne j}}^{3} \lambda [S_+^i S_-^j + S_-^j S_+^i],
\end{equation}
with, $ S_+^i=(\ket{+}\bra{-})_i$, $ S_-^i=(\ket{-}\bra{+})_i$, and $ S_z^i=\frac{1}{2}(\ket{+}\bra{+} - \ket{-}\bra{-})_i$.

In equation (\ref{eq1}), $\omega_i$ is the exciton frequency in each quantum dot $i$, $\Omega_i$ is the excitonic dipoles moments frequency which depends on the external electric field ($\vec{E}$), $\lambda$ denote the Forster interaction \cite{forster1948zwischenmolekulare} i.e. resonant energy transfer between the excitons, and $\hbar J_z$ is the static exciton-exciton dipolar interaction energy. When we introduce into our system thermal fluctuations, the state of the system under thermal equilibrium is given by the reduced density matrix,
\begin{equation}
\label{eq2}
\rho(T)=\frac{1}{Z(T)}  e^{-\beta H}= \sum_{i=1}^{4}P_n(T)\ket{\psi_n}\bra{\psi_n},
\end{equation}
Where $P_n(T)=e^{-\beta E_n}/Z(T)$ are the occupation probabilities of the eigenstates $\ket{\Psi_n}$,
and $Z(T)=\sum_{i=1}^{n}e^{-\beta E_n}$ is the partition function, with $\beta=\frac{1}{K_B T}$, $K_B$ being the Boltzman constant.  

\subsection{Quantum Otto cycle}
In the following section, we describe the quantum Otto engine cycle, operating in four steps.\newline 
\textbf{Step 1}: Quantum isochoric process \cite{quan2007quantum}\newline
The working substance with $\Omega=\Omega_H $ is in the initial state $\rho_1$ and is put in contact with a hot bath at temperature $T=T_H$ until it reaches thermal equilibrium. During this step, the system absorbs an amount of heat from the hot bath ($Q_H>0$). At the end of the process, only the occupation probabilities change to $P_n(T_H)$ while the energy level remains invariant.
\newline
\textbf{Step 2}: Quantum adiabatic process \cite{quan2007quantum,kieu2004second}\newline
The working substance is isolated from the hot bath and the frequency related to the excitonic dipole moment changes from $\Omega_H$ to $\Omega_C$, (with $\Omega_H > \Omega_C$), to satisfy the quantum adiabatic theorem\cite{born1928beweis,messiah1958quantum}. During this step, the system releases an amount of work, without changing the heat.
\newline
\textbf{Step 3}: Quantum isochoric process \newline
The working substance with $\Omega=\Omega_C $ now is in contact with a cold bath at temperature $T=T_C$, and after the thermalization of the system, a quantity of heat is released to the cold bath without a change of work, the occupation probabilities change to $P_n(T_C)$.
\newline
\textbf{Step 4}: Quantum adiabatic process \newline 
The working substance is isolated from the cold bath, and the frequency related to the excitonic dipole moment changes from $\Omega_C$ to $\Omega_H$. During this step, an amount of work is done, but no heat is exchanged. At the end of this step, the working substance returns to the initial condition and is ready for another cycle.
\begin{figure}[H]
	\centering
	\subfloat{\includegraphics[width=0.8\textwidth]{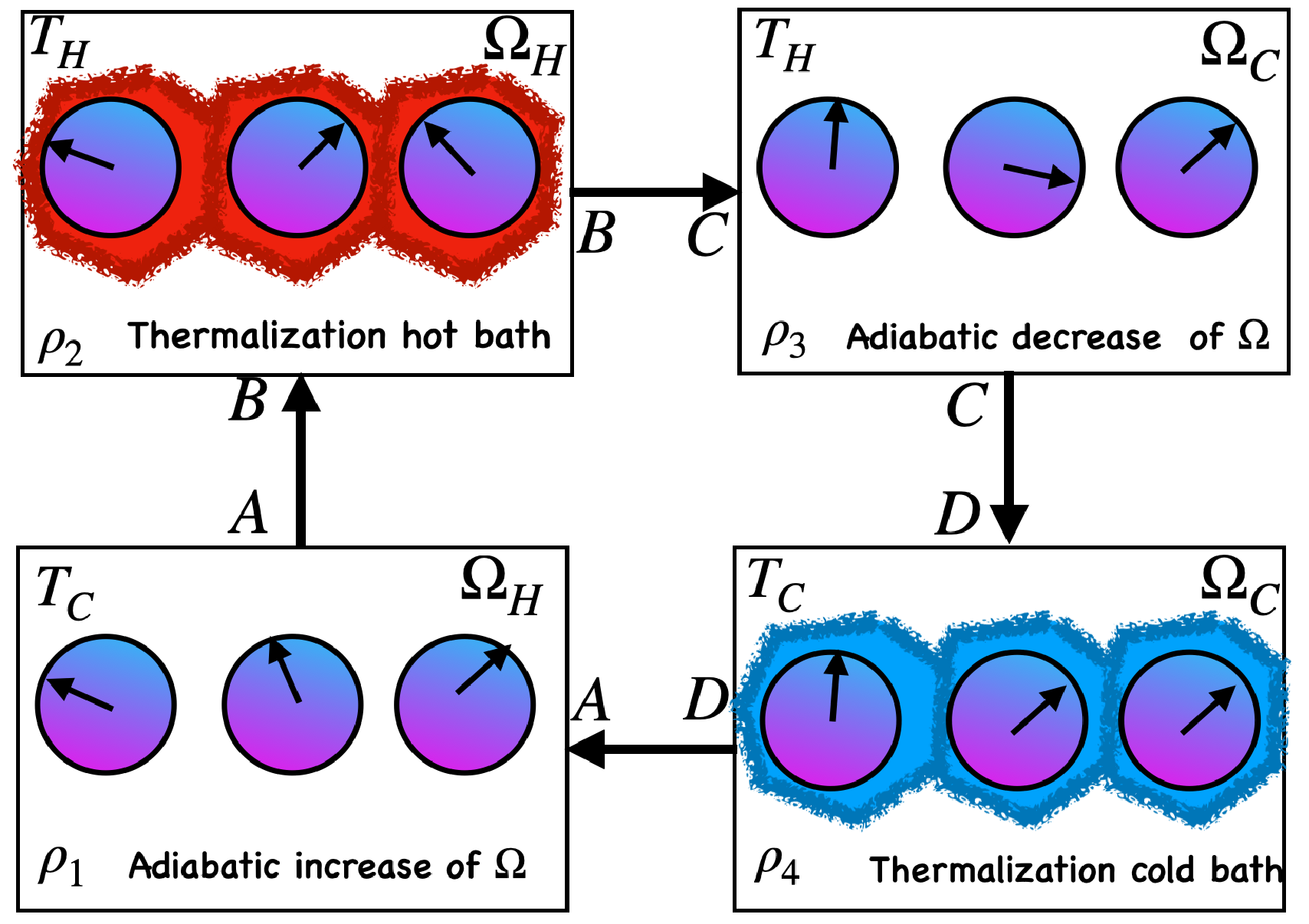}\label{fig0}}
	
	\caption{\label{fig00}} Schematic diagram of the quantum Otto cycle.
\end{figure}
 At each step of the described Otto cycle, the state and the energy of the system are given by,
 \begin{align}
\rho_2=\frac {e^{-\beta_1 H_1}}{Z_1}, \qquad\qquad E_2=Tr \rho_2 H_1,\\
\rho_3=U\rho_2 U^+, \qquad\qquad  E_3=Tr \rho_3 H_2,\\
\rho_4=\frac {e^{-\beta_2 H_2}}{Z_2}, \qquad\qquad E_4=Tr \rho_4 H_2,\\
\rho_1=\Tilde{U}\rho_4 \Tilde{U^+}, \qquad\qquad  E_1=Tr \rho_1 H_1,
\end{align}
where $H_i=H(\Omega_i)$, $Z_i=Tr e^{-\beta_i H_i}$, $i=1,2$, $U$ and $\Tilde{U}$ are unitary operators associated with the second and the fourth steps respectively.

\section{Theory}
\subsection{The Work performed}
In order to describe the work performed and the heat transfer at the quantum level we use the quantum version of the first law of thermodynamics\cite{quan2007quantum,kieu2004second},
\begin{equation}
\label{eq3}
dU=dQ+dW=\sum_n(E_n\,dP_n+P_n \, dE_n)
\end{equation}
In this formula, the heat transfer is related to the change of occupation probabilities, with $dQ=\sum_n E_n\,dP_n$, and the work done is related to the change in energy levels $dW=\sum_n P_n \, dE_n$.\newline 

For the engine described in section 2, the heat absorbed during the first step $Q_H$, the heat liberated during the third step $Q_C$, and $'W'$ the net work performed by the engine are defined by the following expressions:
\begin{equation}
\label{eq4}
Q_{H}=\sum_n E_n^H[P_n(T_H)-P_n(T_L)],
\end{equation}
\begin{equation}
\label{eq5}
Q_{C}=\sum_n E_n^L[P_n(T_L)-P_n(T_H)],
\end{equation}

\begin{equation}
\label{eq6}
W=Q_{H}+Q_{C}=\sum_n[E_n^H-E_n^L][P_n(T_H)-P_n(T_L)],
\end{equation}
The energy levels ($E_n^H$ and $E_n^L$) are obtained by replacing $\Omega$ by $\Omega_H$ and $\Omega_C$ during the first and the third steps respectively. We note that the engine can operate in four modes depending on the signs of W, $Q_H$, and $Q_C$\cite{buffoni2019quantum}:\newline
 $[E]$: $Q_H\ge 0$, $Q_C\le 0$, and $W\ge 0$, in which the machine operates as a quantum heat engine.\newline
 $[R]$: $Q_H\le 0$, $Q_C\ge 0$, and $W\le 0$, in which the machine operates as a refrigerator.\newline
$[A]$: $Q_H\ge 0$, $Q_C\le 0$, and $W\le 0$, in which the machine operates as a thermal accelerator.\newline
$[H]$: $Q_H\le 0$, $Q_C\le 0$, and $W\le 0$, in which the machine operates as a heater.\newline

\subsection{Quantum entanglement}
Quantum entanglement was introduced as the most striking phenomenon in quantum physics with no counterpart in classical physics\cite{einstein1935can}. It is considered a precious resource in several areas of quantum information, quantum computation, quantum communication\cite{nielsen2002quantum}, and their effects in quantum heat engines were established \cite{Huang, Wang,zhang2007,hovhannisyan2013entanglement,brunner2014entanglement,alicki2013entanglement}. It is hard to find an exact expression of this quantity for each quantum system, because the type and the classes of entanglement change with changing the dimension of the quantum systems, and the exact expression is available for a few quantum systems. In this regard, many studies focused on the development of general measures of entanglement. \\
\textbf{Concurrence}\\
For a general two-qubits system, concurrence \cite{PhysRevLett.80.2245} is a frequently used measure of entanglement. It is defined by, 
\begin{equation}
\label{eq_concurrence}
C(\rho)= Max[0,\nu_1-\nu_2-\nu_3-\nu_4],
\end{equation}
where $\nu_i$'s are the square roots of the positive eigenvalues of the matrix $\rho.\Tilde{\rho}$ in decreasing order, with $\Tilde{\rho}$ being a spin flipped of $\rho$, i.e. $\Tilde{\rho}=(\sigma_y \otimes \sigma_y)\rho^*(\sigma_y \otimes \sigma_y)$, 
 $\sigma_y$ and $\rho^*$ being the Pauli matrix and the complex conjugate of $\rho$ respectively.\\
\textbf{Lower bound}\\
For an arbitrary mixed state of three qubits, the lower bound of concurrence \cite{li2009lower} is defined as
\begin{equation}
{\tau}_{3}=\frac{1}{3}\sum_{\alpha}^{6}[(C_{\alpha}^{12/3})^2+(C_{\alpha}^{13/2})^2+(C_{\alpha}^{23/1})^2],
\label{lb}
\end{equation}
with,
\begin{equation}
	C_{\alpha}^{p} = max \{ 0,\lambda(1)_{\alpha}^{p}-\lambda(2)_{\alpha}^{p}
	-\lambda(3)_{\alpha}^{p}-\lambda(4)_{\alpha}^{p}\},
	\label{clb}
\end{equation}
 in which $\lambda(i)_{\alpha}^{p}$ are the square roots of the four nonzero eigenvalues, in decreasing order, of the non-Hermitian matrix $\sqrt{\rho\tilde{\rho}_{\alpha}}$, with $\tilde{\rho}_{\alpha} = (L_{\alpha} \otimes \sigma_y) \rho^{*} (L_{\alpha} \otimes \sigma_y)$, and $L_{\alpha}$ being the generators of $SO(4)$. \\
 
 ${\tau}_{3}$  also characterizes genuine tripartite entanglement that cannot be described by $C(\rho)$ mentioned in \ref{eq_concurrence}, such that, if ${\tau}_{3} =0$, it indicates the absence of any entanglement which means that the state is fully separable. Thus, the lower bound ${\tau}_{3} =0$ can serve as a criterion for detecting and recognizing multipartite entanglement.

\section{Results and discussion} 
In this section, we investigate the behavior of the quantum Otto heat engine, for our system where the energy related to the external electric field $\Vec{E}$ changes between $\hbar \Omega_C$ and $\hbar \Omega_H$  ($\hbar \Omega_H > \hbar \Omega_C$) in the adiabatic processes. Furthermore, we study the effect of the excitons frequency in each quantum dot $\hbar\omega$ and the Forster interaction $\lambda$ on the amount of work performed by the engine. Moreover, we study the relation between the work performed and quantum entanglement between the quantum dot's qubits
\subsection{Work and heat exchange}
\begin{figure}[H]
	\centering
	\subfloat[the work performed, heat liberated $Q_C$ and heat absorbed \\$Q_H$, with $\hbar\omega_i=2meV$, $J_z=2.5 meV$, $T_C=1K$, $T_H=40K$,\\ $\hbar\Omega_H=5meV$, and $\hbar\Omega_C=1meV$.]{\includegraphics[width=0.5\textwidth]{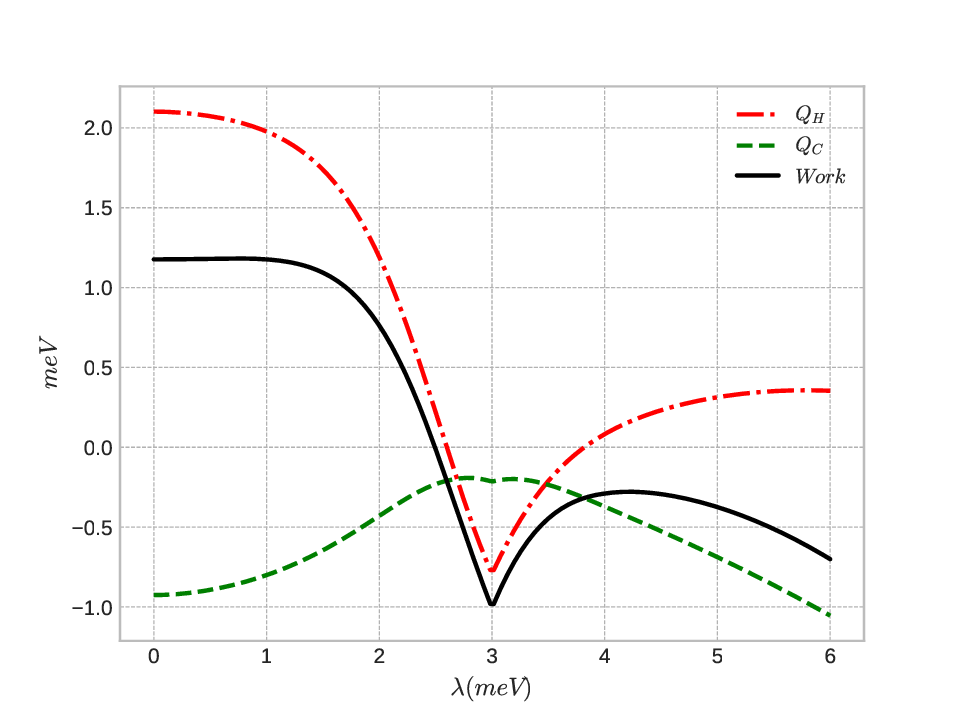}\label{Fig1A}}
\subfloat[The work performed for $\hbar\omega_i=2meV$, $J_z=2.5 meV$, $T_C=1K$, and $\hbar\Omega_C=1meV$.]{\includegraphics[width=0.5\textwidth]{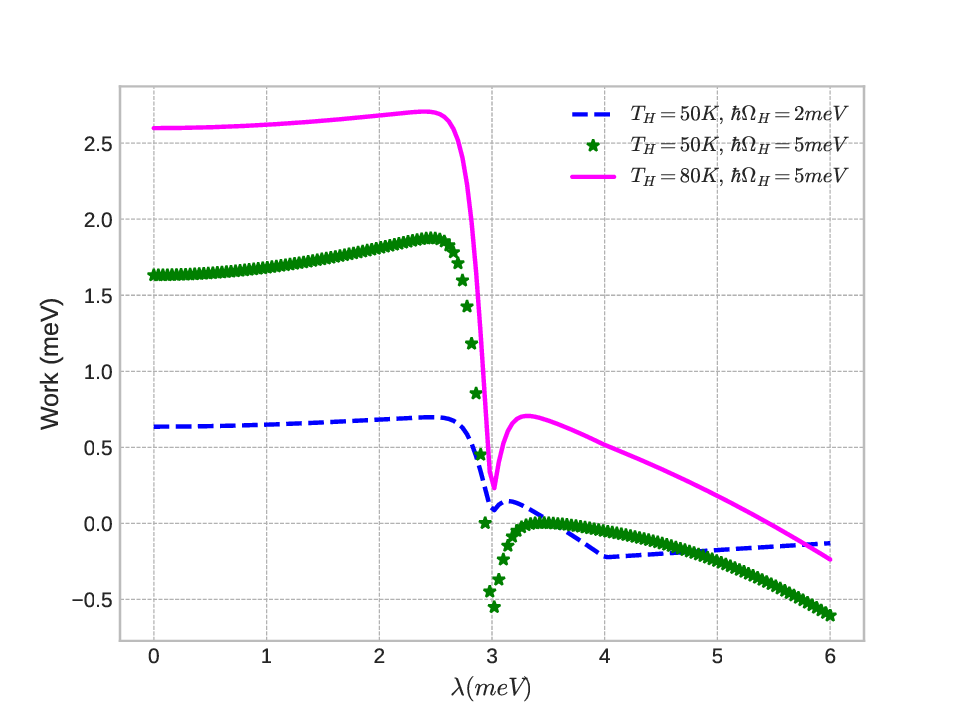}\label{Fig1B}}
 
	\caption{\label{figure1}}
\end{figure}

The behaviors of the work performed, the heat liberated $Q_C$, and the heat absorbed $Q_H$ by the engine against the Forster interaction energy $\lambda$ depicted in Figure \ref{Fig1A}. In figure \ref{Fig1B} we show the effect of the temperature of the hot bath and the energy related to the external electric field $\hbar\Omega_H$ on the work performed by the engine. We note for this model that the energy parameters $\hbar\Omega$, $\hbar\omega$, and $\hbar J_z$ are assumed to be in the order of $meV$ which is coherent with experimental observations \cite{yuan2007exciton,mansour2020quantum}. 

\begin{figure}[H]
	\centering
\subfloat[Variation of Work with $\lambda$ for equal  values of $\hbar \omega_i$; $\omega_1=\omega_2=\omega_2=\omega_3=\omega_i$.]{\includegraphics[width=0.5\textwidth]{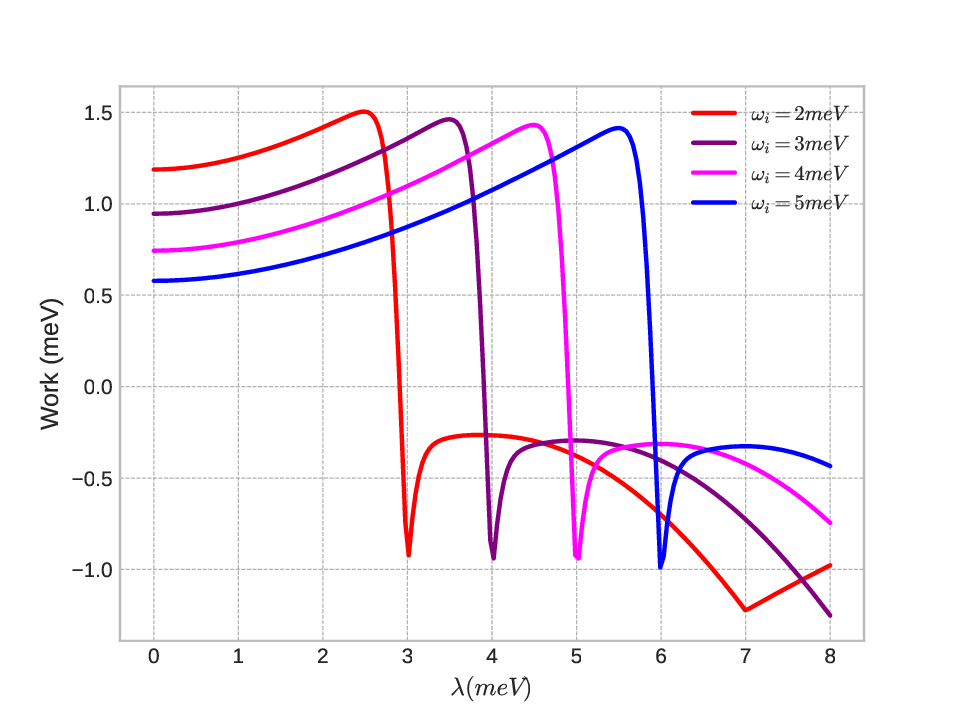}\label{Fig2A}}
 \subfloat[Variation of Work with $\lambda$ for different values of $\hbar \omega_i$ ]{\includegraphics[width=0.5\textwidth]{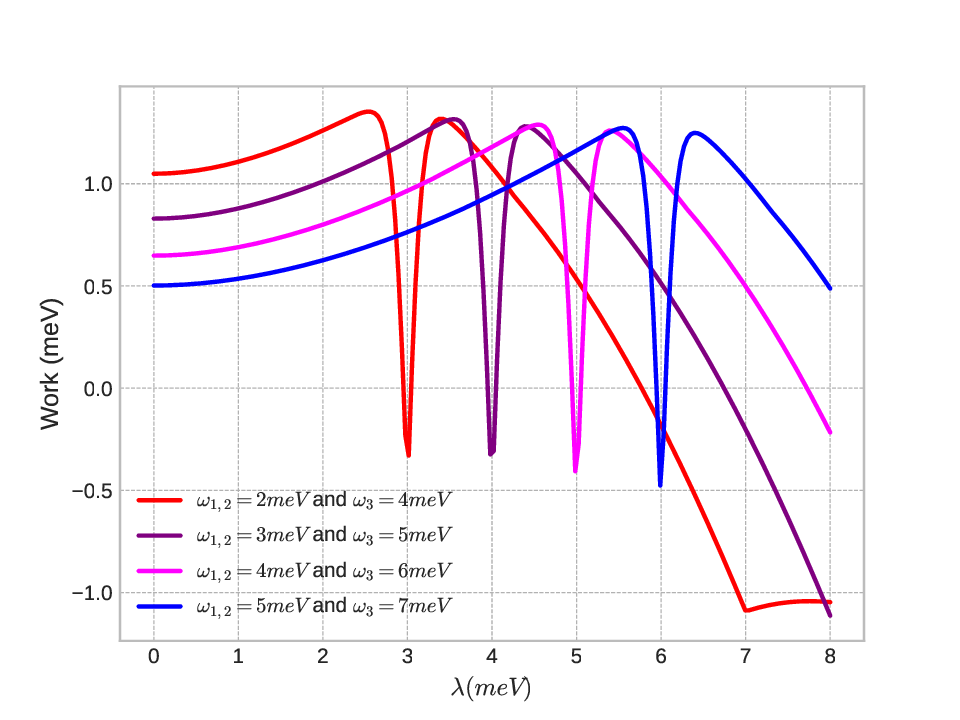}\label{Fig2B}}
	\caption{\label{figure2}}
\end{figure}

Figure \ref{Fig1A} clearly shows distinct regions of the working substance with respect to the Forster interaction energy $\lambda$. In the first region, characterized by smaller values of $\lambda$, positive work is absorbed, indicating that the system functions as a heat engine, this engine absorbs heat from the hot bath $(Q_H>0)$ and transfers it to the cold bath $(Q_C<0)$ while producing useful work. We note that the work is decreased slowly with increasing of the Forster interaction $\lambda$ until a critical value of this parameter after which, the engine requires a negative work ($W<0$) to extract heat from the hot reservoir to the cold reservoir, i.e. the machine works as a heater and the engine cannot produce work. We observe also for a strong Forster interaction, that the heat absorbed $Q_H$ by the engine becomes again positive, but the work and the heat liberated from the engine are still negative. This means that the engine works as a thermal accelerator where the engine transfers heat from the hot reservoir to the cold one, without producing work. These results show that we can switch from one regime to the other by simply increasing a single parameter, the Forster interaction $\lambda$. Figure $\ref{Fig1B}$ shows that the temperature of the hot bath $T_H$ doesn't affect the general behavior (shape) of the work performed but affects only the amount of this work. This amount increases with increasing temperature of the hot bath because the heat absorbed from the hot bath increases. Moreover, the amount of work increases with the excitonic dipoles moments energy $\hbar \Omega$. This is due to the larger difference between $\hbar\Omega_H$ and $\hbar\Omega_C$ under the quantum adiabatic process, which makes the engine convert a lot of energy to work. In order to investigate the effect of the frequency of the excitons in each quantum dot on the work performed by the engine, we present the behaviors of the work performed in figures \ref{Fig2A} and \ref{Fig2B}.
Figure \ref{Fig2A} illustrates the behavior of the work done for different values of the exciton frequency $(\hbar \omega_i)$ as a function of the Foster interaction. In this situation, where the $w_i$'s are equal, we observe, for lower Foster interaction energies, that the work is positive, indicating that the machine operates as a heat engine. However, for stronger Foster interactions, the work becomes negative, rendering the engine incapable of producing any work. Additionally, the critical value of $\lambda$ at which the work turns negative varies with the exciton energy ($\hbar\omega_i$). This suggests that the work is more resilient against the Foster energy for higher exciton energy values, and its behavior remains consistent. Moreover, the amount of the work decreases as $\hbar\omega_i $ increases for smaller values of $\lambda$. Moving on to Figure \ref{Fig2B}, we observe a similar pattern to Figure \ref{Fig2A}. However, in this case, the $\omega_i$ values differ among the three quantum dots. Either two values are the same while the third one is different, or all three values are distinct. Before reaching the critical values of $\lambda$, the work exhibits similar behavior as before. However, as $\lambda$ increases, the work performed undergoes a death and revival, indicating that the engine fails to produce work at certain critical values of $\lambda$. After these critical values, the engine can produce work and can be considered as a heat engine.
Now, the question arises as to why the work done transitions from positive to negative values and why the work exhibits revival when the system lacks symmetry.

\subsection{Work and entanglement}
To understand the reason behind the decrease and subsequent revival of the work at a critical value of $\lambda$ observed in figures \ref{figure1} and \ref{figure2}, we analyze its behavior in comparison to the behavior of several entanglements shared between the quantum dots of the system with different values of the exciton frequency in each quantum dot. We can gain insights into the underlying mechanisms driving this phenomenon by examining the relationship between these two aspects. The decrease and revival of the work refer to a pattern where the work initially declines, then undergoes a resurgence at a specific value of $\lambda$. This behavior can be attributed to the interplay between $\lambda$ and the system's entanglement. By comparing the behavior of the work with the behavior of entanglement, we can observe a correlation between their patterns. The decrease in work is often accompanied by an increase in entanglement, indicating a connection between these phenomena

\begin{figure}[H]
        \centering
\subfloat[$\hbar\omega_1 = \hbar\omega_2 = \hbar\omega_3 =2meV$]{\includegraphics[width=0.5\textwidth]{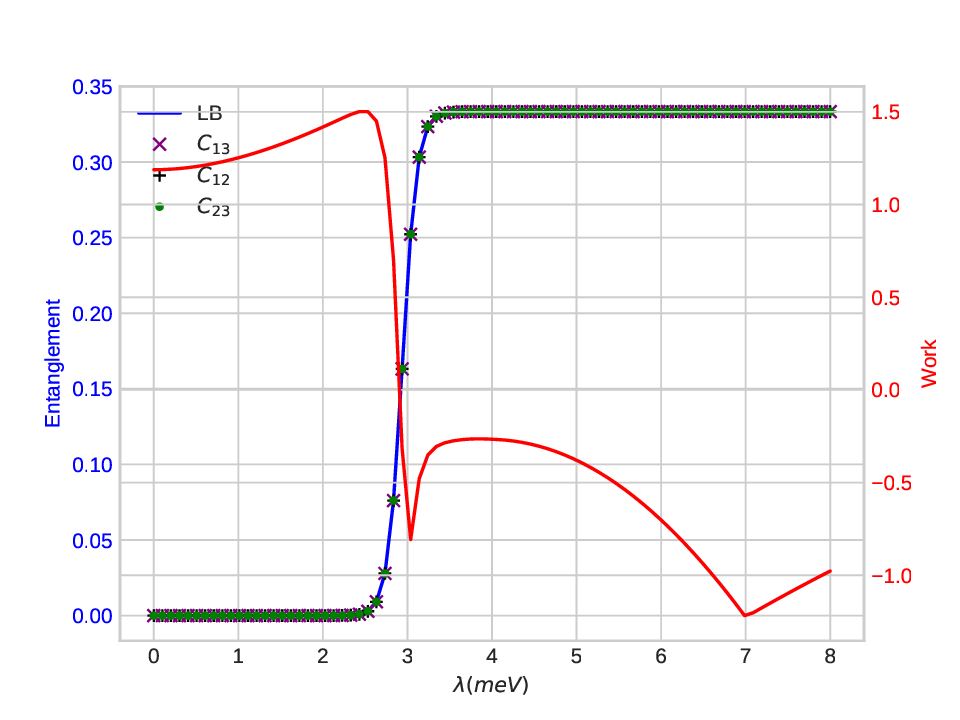}\label{Fig3A}}
\subfloat[$\hbar\omega_1 = \hbar\omega_2 = \hbar\omega_3 =4meV$]{\includegraphics[width=0.5\textwidth]{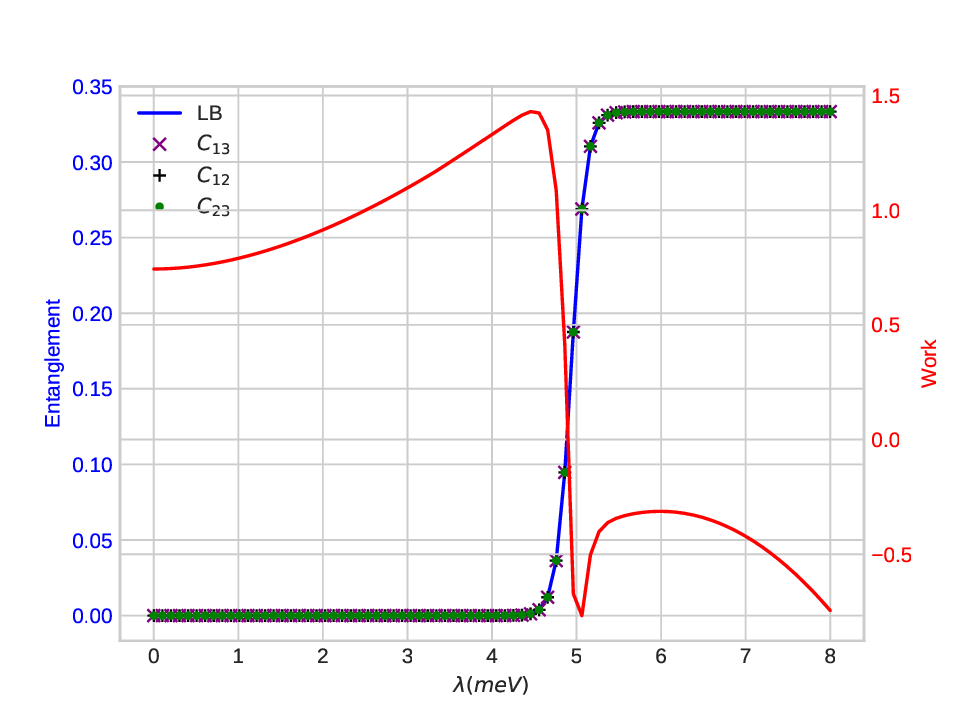}\label{Fig3B}}\\
\subfloat[$\hbar\omega_{1,2} = 2meV$ and  $\hbar\omega_3 =4meV$]{\includegraphics[width=0.5\textwidth]{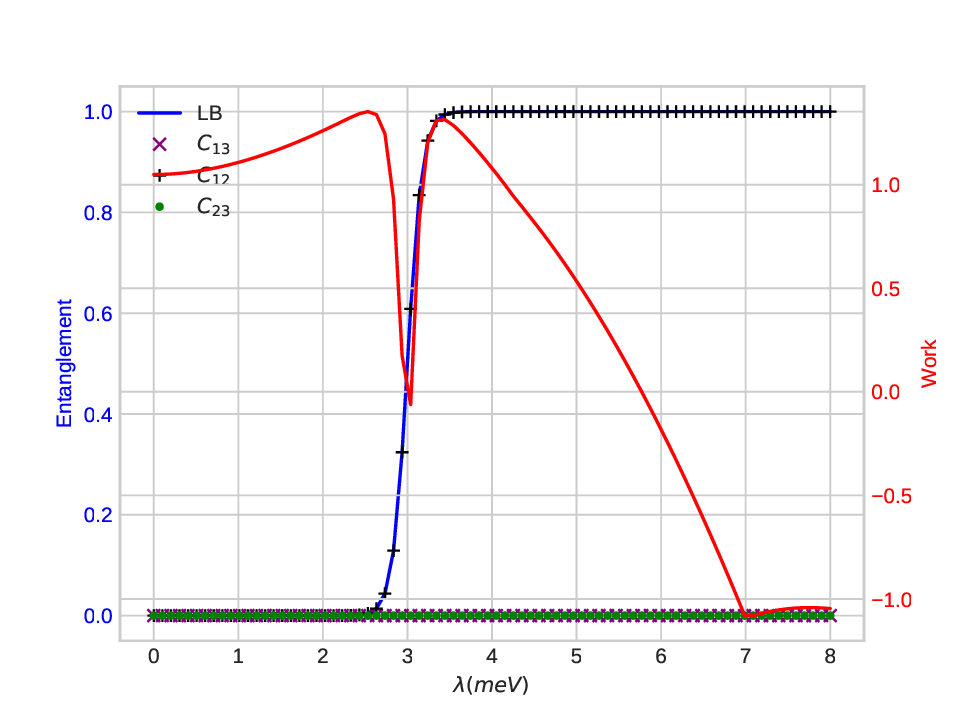}\label{Fig3C}}
\subfloat[$\hbar\omega_{1,2} = 4meV$ and $\hbar\omega_3 =6meV$]{\includegraphics[width=0.5\textwidth]{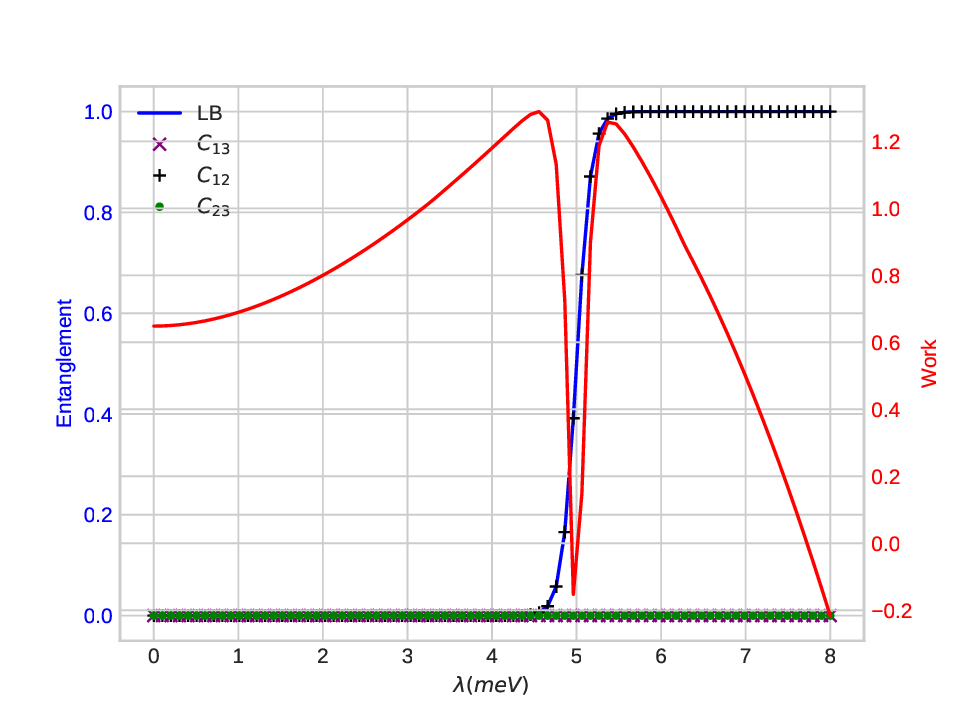}\label{Fig3D}}
	\caption{\label{figure3}Entanglement and Work versus the Förster interaction $\lambda$ for  different values of the exciton frequencies $\hbar\omega_{i}$.}
\end{figure}

It is well known that the entanglement between two quantum systems depends on the strength of the interaction between them. In our case, this interaction can be characterized by a Förster (or dipole-dipole) interaction $\lambda$, which describes the energy transfer between two quantum dots due to their mutual electric dipole moments. The stronger the Förster interaction is the stronger the entanglement between the two quantum dots will be. Indeed, the Förster interaction can lead to the exchange of energy and information between the dots, which can facilitate the establishment of entanglement. However, it is important to note that the presence of the Förster interaction alone does not guarantee the presence of entanglement. Other factors, such as the initial state of the quantum systems and the presence of external noise or decoherence, can also influence the strength of entanglement. Furthermore, in addition to the influence of the exciton frequency in each quantum dot on the critical value of $\lambda$ observed in Figure \ref{figure2}, this frequency plays a significant role in entanglement distribution. The monogamous properties of entanglement dictate that the entanglement between two quantum dots is mutually exclusive. As one quantum dot becomes more entangled with another, its entanglement with other quantum dots diminishes. This monogamous nature of entanglement implies that the frequency of each quantum dot can profoundly impact how entanglement is distributed among the quantum dots in the system.  \\

Figure \ref{figure3} shows that entanglement is absent for small values of $\lambda$  and appears at some critical value  $\lambda_c$, which depends on the exciton frequency in each quantum dot;  $\lambda_c$ becoming larger for larger values of $\hbar\omega$. For example in Figures \ref{Fig3A} and \ref{Fig3B}, we have $\lambda_c=3meV$ and  $\lambda_c=6meV$ for $\hbar\omega_i=2meV$ and $\hbar\omega_i=4meV$ respectively. After $\lambda_c$  we observe that the entanglement increases and reaches its maximum $\frac{1}{3}$. This indicates that the entanglement is evenly distributed among the three quantum dots and is achieved by assigning the same frequency value to each of them. However in Figures \ref{Fig3C} and \ref{Fig3D}  (when $\hbar\omega_{1} = \hbar\omega_{2} \neq \hbar\omega_3$), we observe that only the two quantum dots with identical frequencies exhibit entanglement, reaching its maximum value  $1$ when lambda is larger. This means that these two quantum dots are maximally entangled, while the other pairs remain separated for all values of $\lambda$. Another observation can be drawn is that the critical value of $\lambda$ depends on the frequency of the entangled pairs, and does not depend on the frequency of the other subsystems.\\
Note that at the critical value of $\lambda$, entanglement appears, and the work performed by the engine disappears. However, work exhibits revival when the system lacks symmetry, and entanglement is shared only between two quantum dots. This implies that the revival of work performed after the decline is due to the third quantum dot.\\
Moreover, we can observe that the entanglement at the end of the cycle is detrimental to the work, and the work cannot be obtained if the entanglement is large enough. Our engine cannot produce work if it becomes entangled at the end of its contact with the cold bath. We can conclude that the influence of entanglement on the work performed by a quantum heat engine can be significant, as entanglement can affect the way that energy is distributed within the system. In a quantum heat engine, entanglement can lead to correlations between the energy levels of the system, which can, in turn, affect the efficiency of the engine. For example, if the system is in a highly entangled state, it may be more difficult to extract work from the system, as the energy levels may be more closely correlated and thus more difficult to manipulate. However, if the system is in a less entangled state, it may be easier to extract work from the system, as the energy levels may be less correlated and thus more easily manipulated. Physically, the increase of the excitonic dipoles moments energy $\hbar\Omega$ is due to the increase of the electric field $\vec{E}$, which makes all dipoles parallel, leading to an increase in the repulsive interaction between dipoles, due to the decrease of entanglement between the double quantum dots at the end of the cycle and the increase of the work performed.
Based on the description provided, we can conclude that the system consumes energy in order to create entanglement and then recovers a small quantity after the entanglement is established. This phenomenon can be compared to starting a car, where a relatively significant amount of energy is required to initiate the engine. In this process, various components of the car's system, including the ignition system and fuel pump, are activated. However, certain non-essential functions are temporarily disabled to conserve and concentrate the available energy specifically for starting the engine. This strategic approach ensures that the system can channel its resources effectively and concentrate its energy on the primary objective of entanglement creation; this can lead to an entanglement cost, approach study, based on these types of heat engines.

\section{Conclusion}
This paper delved into the intricate world of quantum heat engines, focusing on the quantum Otto cycle with the working substance consisting of three InAs quantum dots. These engines can operate at extremely small scales, offering promising prospects for efficient energy conversion. We have studied their behavior under the influence of external electric field and Forster interaction parameters, shedding light on the fascinating interplay between these parameters and the engine's performance. We find that the work performed by the quantum heat engine is strongly influenced by the Forster interaction energy ($\lambda$) and the temperature of the baths. We have identified distinct regions of operation based on the Forster energy,  ranging from a heat engine that absorbs heat and produces work to a thermal accelerator that transfers heat without producing work. Temperature variations in the hot bath directly impact the amount of work extracted, with higher temperatures leading to increased work production. Additionally, the energy of excitonic dipoles ($\hbar \Omega$) plays a significant role, with larger differences between $\hbar \Omega_H$ and $\hbar \Omega_C$ yielding more work.\\
The study also explored the connection between the work performed by the engine and quantum entanglement among the quantum dots. As the Forster interaction increased, work initially decreased and then revived at critical values of $\lambda$. This pattern is closely correlated with changes in quantum entanglement. Higher entanglement was associated with decreased work, emphasizing the significance of entanglement in influencing engine performance. 
Then, the work highlighted that the presence of the Förster interaction contributes to entanglement between quantum dots, which can affect the energy distribution and the efficiency of the engine. The entanglement properties are intricately related to the exciton frequency in each quantum dot, with certain frequency configurations favoring maximal entanglement. Importantly, the presence of entanglement at the end of the cycle can inhibit work production, demonstrating the substantial impact of entanglement on quantum heat engine behavior. Finally, The study suggests that the system incurs an energy cost to establish entanglement. Resources are strategically allocated to initiate the entanglement process, with non-essential functions temporarily disabled. This approach ensures efficient energy utilization, leading to the establishment of entanglement.

In summary, this research unveils the intricate relationship between the Forster interaction, entanglement, and the performance of quantum heat engines. The findings underscore the potential of quantum heat engines as a viable energy conversion technology, while also highlighting the importance of managing entanglement in optimizing their efficiency. Further exploration of these phenomena could pave the way for the development of highly efficient quantum heat engines, contributing to the ever-evolving landscape of energy technology.

\bibliographystyle{IEEEtran}
\bibliography{sample}

\end{document}